\documentclass[prb,twocolumn,showpacs,preprintnumbers,amsmath,amssymb]{revtex4}
\usepackage{mathrsfs}
\usepackage{graphicx}

\begin{document}

\title{Diamagnetism versus Paramagnetism in charged spin-1 Bose gases }

\author{Xiaoling Jian}
\author{Jihong Qin}
\author{Qiang Gu}\email[Corresponding author: ]{qgu@ustb.edu.cn}
\affiliation{Department of Physics, University of Science and
Technology Beijing, Beijing 100083, China}

\date{\today}

\begin{abstract}

It has been suggested that either diamagnetism or paramagnetism of
Bose gases, due to the charge or spin degrees of freedom
respectively, appears solely to be extraordinarily strong. We
investigate magnetic properties of charged spin-1 Bose gases in
external magnetic field, focusing on the competition between the
diamagnetism and paramagnetism, using the Lande-factor $g$ of
particles to evaluate the strength of paramagnetic effect. We
propose that a gas with $g<{1/\sqrt{8}}$ exhibits diamagnetism at
all temperatures, while a gas with $g>{1/2}$ always exhibits
paramagnetism. Moreover, a gas with the Lande-factor in between
shows a shift from paramagnetism to diamagnetism as the temperature
decreases. The paramagnetic and diamagnetic contributions to the
total magnetization density are also calculated in order to
demonstrate some details of the competition.
\end{abstract}

\pacs{05.30.Jp, 75.20.-g, 75.10.Lp, 74.20.Mn}

\maketitle

\section{Introduction}

The Bose gas plays a significant role in understanding a series of
exotic quantum phenomena, including superfluidity and
superconductivity. It is well known that the ideal Bose gas exhibits
the Bose-Einstein condensation (BEC) below a critical temperature.
In 1938, London first connected BEC to the $\lambda$ transition in
Helium.\cite{London1938} In 1946, Ogg proposed that the BEC of
bosonic electron-pairs might result in
superconductivity.\cite{Ogg1949} Furthermore,
Schafroth\cite{Schafroth1955} and Blatt and Butler\cite{Blatt1955}
showed that an ideal gas of charged bosons exhibits the essential
equilibrium features of a superconductor. Although the BCS
theory\cite{BCS1957} revealed that electrons in a superconductor
form Cooper-pairs, as opposed to real-space pairs, the
Schafroth-Blatt-Butler theory is helpful to the understanding of
superconductivity and more importantly it stimulated research
interest in charged Bose gases.

The charged Bose gas (CBG) is solely of academic interest in its own
right. Especially, it exhibits nontrivial magnetic properties. For
example, the condensation phenomenon in CBGs is strongly affected by
the external magnetic field, owing to the quantization of the
orbital motion of charged particles in magnetic
field.\cite{Osborne1949,Schafroth1955,May1965,Arias1989,Alexandrov1993,Daicic,Toms,Rojas1996}
Schafroth clearly indicated that an arbitrarily small value of the
magnetic field introduces qualitative changes: BEC does no longer
occur.\cite{Schafroth1955} May extended this idea to a
$d$-dimensional CBG and found that it can condense only for $d >
4$.\cite{May1965} This point was reexamined by other researchers
based on different methods.\cite{Daicic,Toms} Meanwhile, the orbital
motion results in extremely large Landau diamagnetism in CBGs. It
was pointed out that the 3-dimensional CBG displays Meissner effect
at low temperatures.\cite{Schafroth1955,Rojas1996,Toms1997} More
recently, Alexandrov quantitatively accounted for the enhanced
normal-state diamagnetism of superconducting cuprates using the CBG
model.\cite{Alexandrov2006}

On the other hand, neutral bosons with spin (spinor bosons) have
also been studied theoretically.\cite{Yamada1982} The realization of
spinor Bose condensate in optically trapped alkali
atoms\cite{Stamper-Kurn1998} stimulates new research interest. The
constituent atoms, such as $^{87}\text{Rb}$, $^{23}\text{Na}$, and
$^{7}\text{Li}$ have (hyperfine) spin degrees of freedom and thus a
magnetic moment. Their spin degrees of freedom become active in
purely optical traps and thus investigation of their magnetic
properties becomes possible. Yamada\cite{Yamada1982}, and Simkin and
Cohen\cite{Simkin1999} calculated the magnetization of neutral
spinor bosons in magnetic field and found that once BEC takes place,
the magnetization remains finite even if $H=0$, as if the system was
magnetized spontaneously. The zero-field susceptibility tends to
diverge as the temperature goes down to the BEC temperature.
Moreover, rigorous proofs presented by Eisenberg and Lieb show that
the magnetization and zero-field susceptibility at finite
temperatures exceed that of a pure paramagnet.\cite{Eisenberg2002}
All the results indicate that the neutral spinor bosons take on
extraordinary paramagnetic effect in magnetic field.

Bearing in mind the two contrary features of the Bose gas, one
immediate question which must be raised is: which kind of magnetism
will manifest itself if the bosons possess both the charge and spin
degrees of freedom? Analogously, this question has been answered for
charged Fermi gases, e.g., the electron gas, where both the
paramagnetism and diamagnetism are relatively weak. For the electron
gas, the diamagnetic part of the zero-field susceptibility is
one-third (in absolute value) of the paramagnetic part, so
altogether the gas is paramagnetic. For charged spinor Bose gas,
since both the paramagnetism and diamagnetism can be extremely
large, their competition is a more interesting problem.

In this paper, we study the competition between paramagnetism and
diamagnetism of a charged spin-1 Bose gas in external magnetic
field. In Section II, a model consisting of both the Landau
diamagnetic effect and Pauli paramagnetic effect is proposed. The
total magnetization density as well as its paramagnetic and
diamagnetic parts are calculated respectively. Section III presents
a detailed discussion of the obtained results. In Section IV, a
brief summary is given.

\section{The model}

The orbital motion of a charged boson with charge $q$ and mass $m^*$
in a constant magnetic field $B$ is quantized into the Landau
levels,
\begin{eqnarray}\label{Diam}
\epsilon^l_{jk_{z}}=(\frac{1}{2}+j)\hbar\omega + \frac{\hbar^{2}
k_{z}^{2}}{2m^{\ast}},
\end{eqnarray}
where $j=0,1,2,\ldots$ labels different Laudau levels and $\omega
=qB/(m^{\ast}c)$ is the gyromagnetic frequency. Since
$\omega\varpropto B$, $\omega$ can be used to indicate the magnitude
of the magnetic field in the following discussions. We assume the
magnetic field is in the $z$ direction. Each Landau level is
degenerate with degeneracy equal to
\begin{eqnarray}\label{g}
D_L=\frac{qBL_{x}L_{y}}{2\pi\hbar c}.
\end{eqnarray}
Here we suppose the gas is in a box with $L_{i}\rightarrow \infty$,
where $L_{i}$ is the length of the box in the $i$th direction. For a
spin-1 boson, the Zeeman energy levels split in the magnetic field
due to the intrinsic magnetic moment associated with the spin degree
of freedom,
\begin{eqnarray}\label{Param}
\epsilon_{\sigma}^{ze} =  -  g\frac {\hbar q}{m^*c} \sigma B,
\end{eqnarray}
where $\sigma$ refers to the spin-$z$ index of Zeeman state $\left|
{F=1,m_F=\sigma} \right\rangle$ ($\sigma= +1, 0, -1$) and $g$ is the
Lande-factor. The quantization of the orbital motion and the Zeeman
effect give rise to the Landau diamagnetism and Pauli paramagnetism,
respectively.

We consider an assembly of $N$ bosons, whose effective Hamiltonian
reads
\begin{eqnarray}\label{Hamil}
\bar{H} - \mu {N}= D_L\sum_{j,k_z,\sigma}\left(\epsilon^l_{jk_{z}}
+\epsilon_{\sigma}^{ze}  - \mu\right)n_{jk_z\sigma},
\end{eqnarray}
where $\mu$ is the chemical potential. The charged spinor bosons
have been discussed theoretically in the context of relativistic
pair creation\cite{Daicic1995}. However, magnetism of charged spinor
Bose gases is less studied and of major interest in the present
work.

The grand thermodynamic potential is formally expressed as
\begin{align}\label{T1}
\Omega_{T\neq0}& =-\frac{1}{\beta}\ln\mathrm{Tr}e^{-\beta(\bar{H}-\mu N)}\nonumber \\
&=\frac{1}{\beta}D_L\sum_{j,k_z,\sigma}\ln
[1-e^{-\beta(\epsilon^l_{jk_{z}} +\epsilon_{\sigma}^{ze} - \mu )}],
\end{align}
where $\beta=(k_{B}T)^{-1}$. Converting the sum over $k_z$ to
continuum integral, we get
\begin{align}\label{T3}
\Omega_{T\neq0}=&\frac{\omega
m^{\ast}V}{(2\pi)^{2}\hbar\beta}\sum_{j=0}^{\infty}\sum_{\sigma}\int dk_z\nonumber\\
&\times\ln\{1-e^{-\beta[(j+\frac{1}{2})\hbar\omega+\frac{\hbar^{2}k_z^{2}}{2m^{\ast}}
 -g\frac {\hbar q}{m^*c} \sigma B-\mu]}\},
\end{align}
where $V$ is the volume of the system. Then using the Taylor
expansion and performing the integral over $k_z$, we have
\begin{align}\label{T2}
\Omega_{T\neq0}=&-\frac{\omega
V}{\hbar^{2}}\left(\frac{m^{\ast}}{2\pi\beta}\right)^{3/2}\nonumber\\
&\times\sum_{l=1}^{\infty}\sum_{\sigma}\frac{l^{-\frac{3}{2}}e^{-l\beta(\frac{\hbar\omega}{2}
 -g\frac{\hbar q}{m^{\ast}c} \sigma
B- \mu)}}{1-e^{-l\beta\hbar\omega}}.
\end{align}
This treatment has been used by Standen and Toms\cite{Toms} to deal
with scalar Bose gases. As they mentioned, this theory is more
reliable at high temperature. Similarly, we introduce some compact
notation for the class of sums,
\begin{align}\label{s}
\Sigma_{\kappa}[\alpha,\delta]=\sum_{l=1}^{\infty}\frac{l^{\alpha/2}e^{-lx(\varepsilon+\delta)}}{(1-e^{-lx})^{\kappa}},
\end{align}
where $x=\beta\hbar\omega$ and $\mu+g\hbar q\sigma B/(m^{\ast}c)
=(\frac{1}{2}-\varepsilon)\hbar\omega$. With this notation we may
rewrite Eq.~(\ref{T2}) as
\begin{align}\label{Ts}
\Omega_{T\neq0}=-\frac{\omega
V}{\hbar^{2}}\left(\frac{m^{\ast}}{2\pi\beta}\right)^{3/2}\sum_{\sigma}\Sigma_{1}[-D,0],
\end{align}
where $D=3$. Then the density of particles $n=N/V$ can be derived
from the thermodynamic potential,
\begin{align}\label{n}
n &=-\frac{1}{V}\left(\frac{\partial\Omega}{\partial\mu}\right)_{T,V}\nonumber\\
&=x\left(\frac{m^{\ast}}{2\pi\beta\hbar^{2}}\right)^{3/2}\sum_{\sigma}\Sigma_{1}[2-D,0].
\end{align}
The magnetization density $M$ is written as
\begin{align}\label{Mag}
M_{T\neq 0} &=-\frac{1}{V}\left(\frac{\partial\Omega}{\partial B}\right)_{T,V}\nonumber\\
&=\frac{\hbar q}{m^{\ast}c}\left(\frac{m^{\ast}}{2\pi\beta\hbar^{2}}\right)^{3/2}\sum_{\sigma}\biggl\{\Sigma_{1}[-D,0] \nonumber\\
&+x(g\sigma-\frac{1}{2})
\Sigma_{1}[2-D,0]-x\Sigma_{2}[2-D,1]\biggr\}.
\end{align}
It is convenient to introduce some dimensionless parameters, such as
$\overline{M}=m^{\ast}cM/(n\hbar q)$,
$\overline{\omega}=\hbar\omega/(k_{B}T^{\ast})$, $t=T/T^{\ast}$ and
$x^{\prime}=\overline{\omega}/t$, to re-express equations (\ref{n})
and (\ref{Mag}),
\begin{align}\label{n1}
1=\overline{\omega}t^{1/2}\sum_{\sigma}\Sigma_{1}^{\prime}[2-D,0],
\end{align}
\begin{align}\label{M1}
\overline{M}_{T\neq 0}
=&\;t^{3/2}\sum_{\sigma}\biggl\{\Sigma_{1}^{\prime}[-D,0]+x^{\prime}(g\sigma-\frac{1}{2})\nonumber\\
&\times\Sigma_{1}^{\prime}[2-D,0]-x^{\prime}\Sigma_{2}^{\prime}[2-D,1]\biggr\}.
\end{align}
Here the characteristic temperature $T^{\ast}$ is given by
$k_{B}T^{\ast}=2\pi\hbar^{2} n^{2/3}/m^{\ast}$. The Bose-Einstein
condensation temperature of spin-1 Bose gas with density $n$ is just
defined as $k_{B}T_c = 2\pi\hbar^{2}
n^{2/3}/\{m^{\ast}[3\zeta(3/2)]^{2/3}\}\approx k_{B}T^{\ast}/3.945$.
The dimensionless notation $\Sigma_{\kappa}^{\prime}[\alpha,\delta]$
should be
\begin{align}\label{s1}
\Sigma_{\kappa}^{\prime}[\alpha,\delta]=\sum_{l=1}^{\infty}\frac{l^{\alpha/2}e^{-lx^{\prime}(\varepsilon+\delta)}}{(1-e^{-lx^{\prime}})^{\kappa}},
\end{align}
where $\mu^{\prime}=\mu/(k_{B}T^{\ast})$ and $\mu^{\prime}+g\sigma
\overline{\omega}=(\frac{1}{2}-\varepsilon)\overline{\omega}$. The
two variables $\mu^{\prime}$ and $\overline{M}$ are determined by
Eqs. (\ref{n1}) and (\ref{M1}).

\section{Results and discussions}

\begin {figure}[t]
\includegraphics[width=0.45\textwidth,keepaspectratio=true]{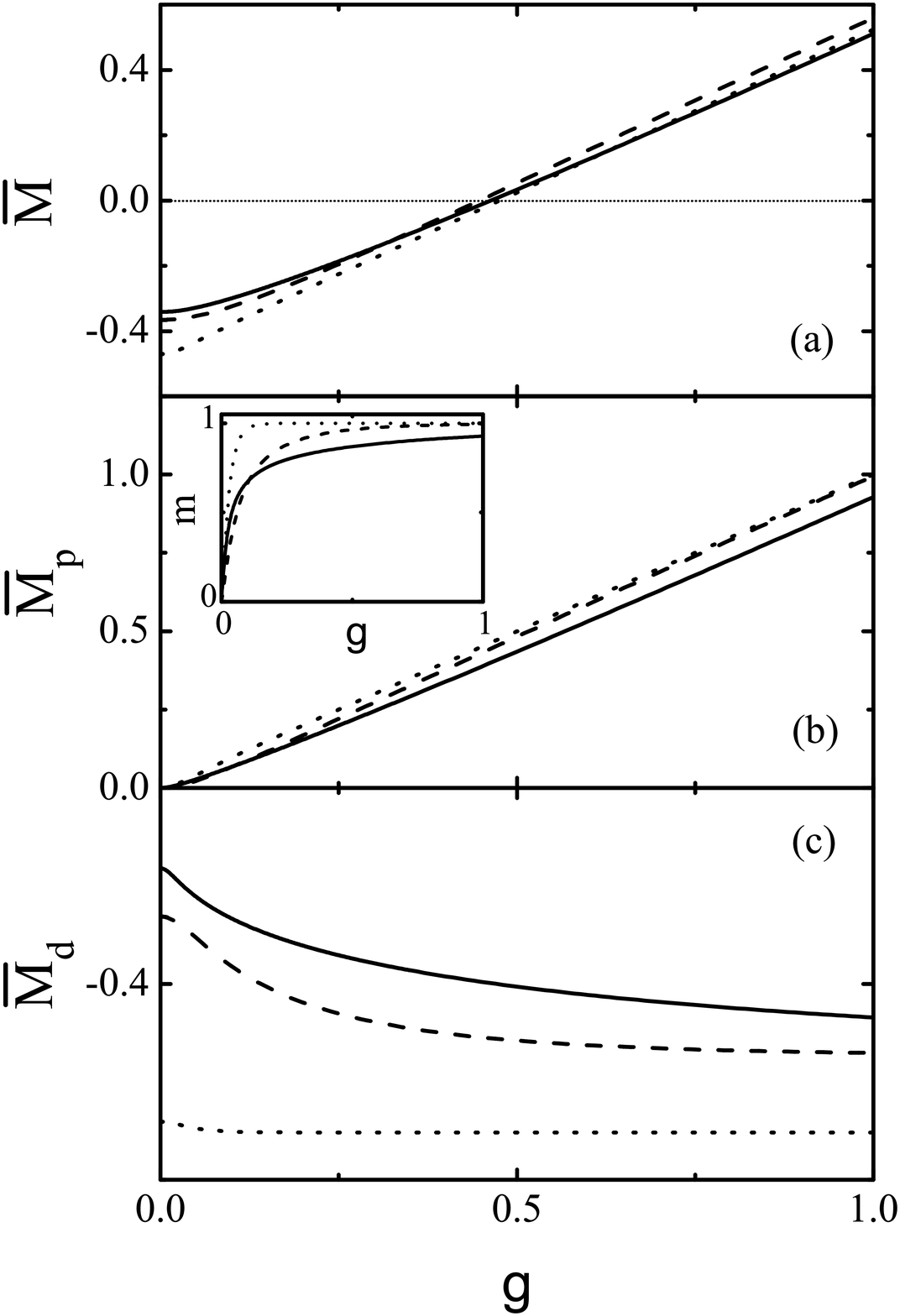}
\caption{(a) The total magnetization density ($\overline{M}$), (b)
the paramagnetization density ($\overline{M}_{p}$), and (c) the
diamagnetization density ($\overline{M}_{d}$) as a function of $g$
for fixed magnetic fields at $t=0.1$. Here the solid line, dashed
line and dotted line correspond to $\overline{\omega}= 0.05$, $0.3$,
and $3$, respectively. Inset: $m$ as a function of $g$.
 }\label{M-G}
\end{figure}

In our model, both the charge and spin degrees of freedom are taken
into account, which are described respectively by the Landau energy
term $\epsilon^l_{jk_{z}}$ and the Zeeman energy term
$\epsilon_{\sigma}^{ze}$ in the Hamiltonian (\ref{Hamil}). In the
case that the Lande-factor $g$ tends to zero, the model degenerates
into a charged scalar boson model which exhibits strong diamagnetism
as already intensively
discussed.\cite{May1965,Arias1989,Alexandrov1993,Daicic,Toms,Rojas1996}
As $g$ becomes larger, the paramagnetic effect is strengthened.

\begin {figure}[t]
\includegraphics[width=0.45\textwidth,keepaspectratio=true]{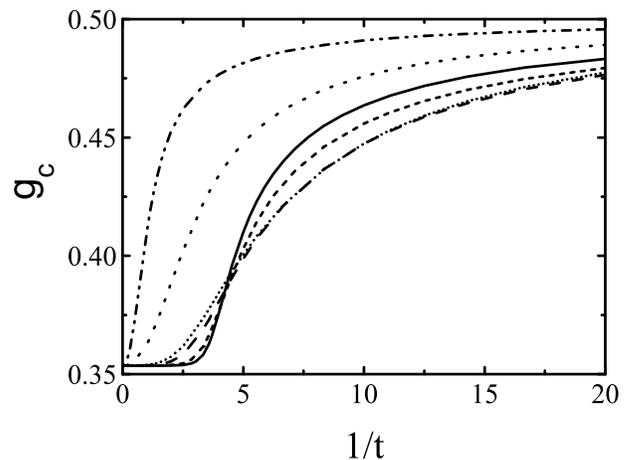}
\caption{Plots of the critical value of Lande-factor, $g_{c}$ as a
function of $1/t$ for fixed values of $\overline{\omega}$. The field
is chosen as: $\overline{\omega}$ = 10 (dash-dot-dotted line), 3
(dotted line), 0.5 (short dotted line), 0.3 (dashed line), 0.1
(short-dashed line), and 0.05 (solid line).} \label{Gc}
\end{figure}

We calculate the dimensionless magnetization density $\overline{M}$
as a function of $g$, as shown in Fig. \ref{M-G}(a). $\overline{M}$
is negative in the small $g$ region, which means that the
diamagnetism dominates. The absolute value of $\overline{M}$ is
larger in the stronger field $\overline{\omega}$. For each given
value of $\overline{\omega}$, $\overline{M}$ increases monotonically
with $g$. $\overline{M}$ changes its sign from negative to positive
at a critical value of $g$, noted as $g_c$ hereinafter, reflecting
that the paramagnetism becomes dominant as $g$ increases. Note that
the slope of the $\overline{M}$ curve is dependent on $g$. When $g$
is near to zero, $\overline{M}$ increases slowly but the slope rises
quickly with $g$. It means that the interplay between diamagnetism
and paramagnetism is complex and nonlinear. However, in the strong
paramagnetic region, $\overline{M}$ grows almost linearly with $g$.

Figure \ref{M-G}(b) plots the paramagnetic contribution to
$\overline{M}$ (named as the paramagnetization density),
$\overline{M}_{p}= g m$, with an inset showing $m= n_{1}-n_{-1}$,
and Figure \ref{M-G}(c) shows the diamagnetic contribution to
$\overline{M}$ (named as the diamagnetization density),
$\overline{M}_{d}= \overline{M}-\overline{M}_p$. The increasing
tendency of $\overline{M}_{p}$ is similar to that of $\overline{M}$.
It is noteworthy that the diamagnetization density is not
suppressed, but enhanced as $g$ becomes larger. Comparing Figs
\ref{M-G}(a), \ref{M-G}(b) and \ref{M-G}(c), it can be seen that the
increase in $\overline{M}$ comes from the paramagnetic effect. In
the small $g$ region, both $\overline{M}_{p}$ and $\overline{M}_{d}$
are strengthened nonlinearly with increasing $g$. As shown in the
inset of Fig. \ref{M-G}(b), $m$ grows very quickly. Nevertheless,
both $m$ and $\overline{M}_{d}$ flatten out in the large $g$ region.
So the slope of $\overline{M}$ curve is mainly due to the
paramagnetization.

\begin {figure}[t]
\includegraphics[width=0.45\textwidth,keepaspectratio=true]{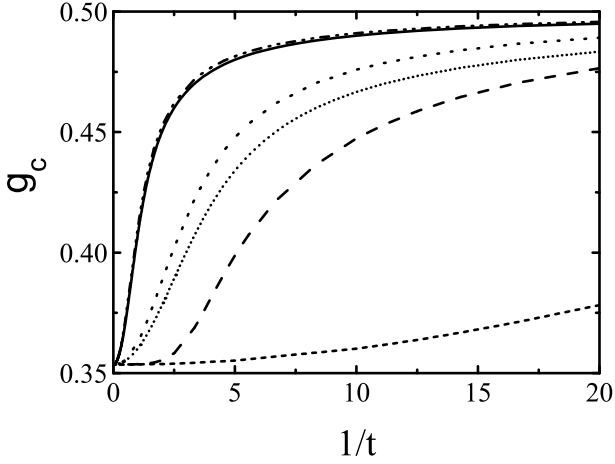}
\caption{Curves of $g_{c}- 1/t$ obtained respectively from the
Bose-Einstein (BE) and Maxwell-Boltzmann (MB) statistics with fixed
values of $\overline{\omega}$. Where $\overline{\omega}$ = 10
(dash-dot-dotted line, BE; solid line, MB), 3 ( dotted line, BE;
short dotted line, MB), and 0.3 (dashed line, BE; short dashed line,
MB).} \label{Gc-Bz}
\end{figure}

According to discussions above, the critical value of the
Lande-factor, $g_{c}$, is an important parameter to describe the
competition between the diamagnetism and paramagnetism. $g_{c}$ is a
function of the temperature $t$ and the magnetic field
$\overline{\omega}$. Figure \ref{Gc} shows $g_{c}$ as a function of
$1/t$ in different magnetic fields $\overline{\omega}=10$, 3, 0.5,
0.3, 0.1 and 0.05. Obviously, $g_{c}$ varies monotonically with the
temperature $t$, while its dependence on the field
$\overline{\omega}$ is not simple. For example, in the low
temperature region, $g_{c}$ decreases with decreasing
$\overline{\omega}$ at a given temperature as $\overline{\omega}$ is
still larger than $0.3$, then it rises up as $\overline{\omega}$
goes down further from $\overline{\omega}\approx 0.3$.

As already mentioned, the results of our theory are more credible at
high temperature. In the high temperature limit, $g_{c}$ seems
universal with respect to different choices of magnetic field,
$g_{c}|_{t\to\infty}\approx 0.35356$. For a given magnetic field,
$g_{c}$ increases as the temperature falls down. This suggests that
the diamagnetic region is larger at low temperatures than at high
temperatures.  Although the exact value of $g_{c}$ can not be
obtained at very low temperatures, its variation trend can be
estimated from Fig.~\ref{Gc}. It seems that $g_{c}$ ranges from
$0.475$ to $0.50$ in various magnetic fields.

\begin {figure}[t]
\includegraphics[width=0.45\textwidth,keepaspectratio=true]{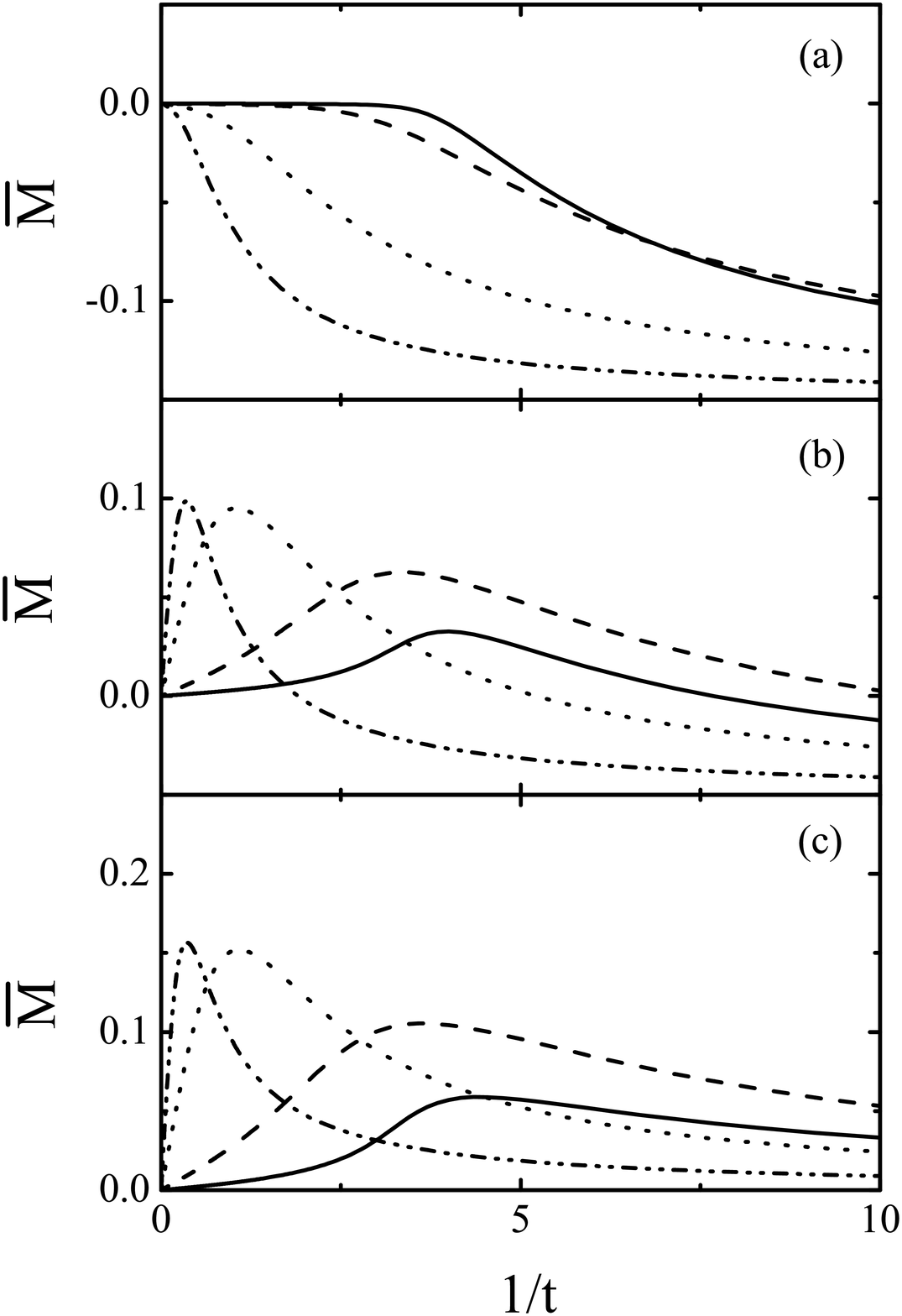}
\caption{Shown are plots of dimensionless $\overline{M}$ as a
function of $1/t$ for each given value of $\overline{\omega}$ at a
fixed $g$. The value of $\overline{\omega}$ were in sequence as 10
(dash-dot-dotted line), 3 (dotted line), 0.3 (dashed line), and 0.05
(solid line). (a) corresponds to $g=0.35$. (b) corresponds to
$g=0.45$. (c) corresponds to $g=0.5$.} \label{M}
\end{figure}

\begin {figure*}[t]
\includegraphics[width=0.95\textwidth,keepaspectratio=true]{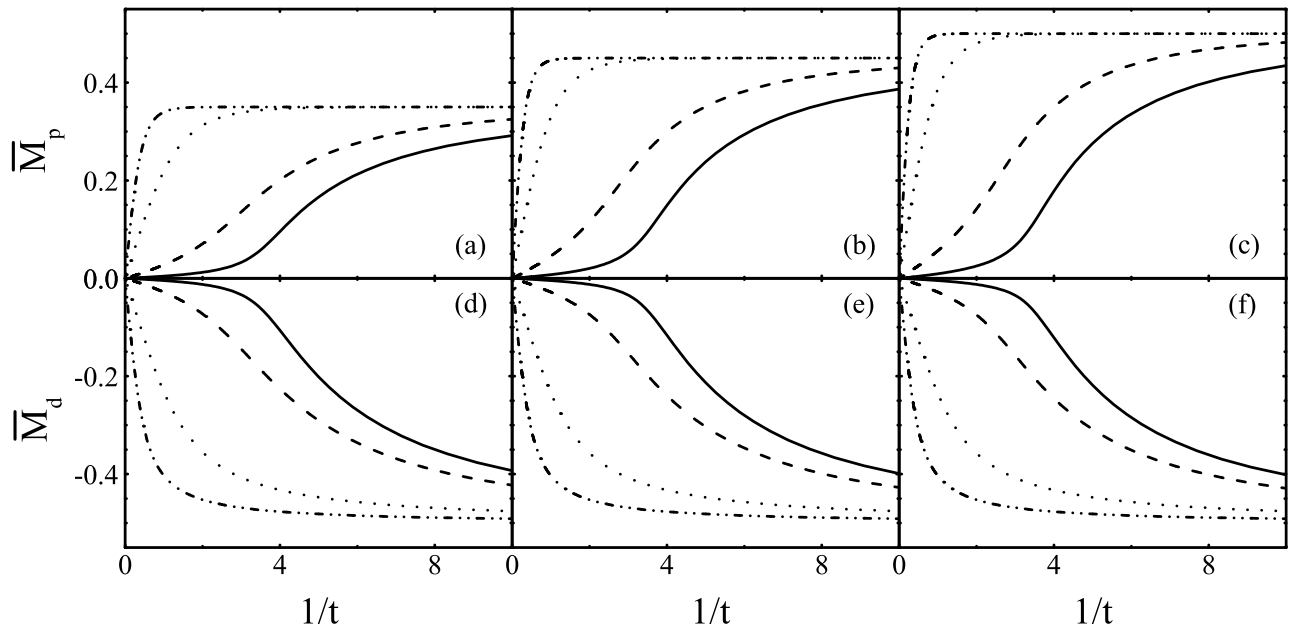}
\caption{The dimensionless $\overline{M}_{p}$ and $\overline{M}_{d}$
as a function of $1/t$ for fixed values of $\overline{\omega}$. From
left to right $g$ = 0.35, 0.45 and 0.5, respectively. For each given
value of $g$, the field is chosen as: $\overline{\omega}$ = 10
(dash-dot-dotted line), 3 (dotted line), 0.3 (dashed line), and 0.05
(solid line), respectively.} \label{Mt}
\end{figure*}

It is useful to reexamine the high temperature behaviors of $g_{c}$
by generalizing above calculations to a spin-1 Boltzmann gas, since
the Bose-Einstein statistics reduces to Maxwell-Boltzmann statistics
in the high temperature limit. The grand thermodynamic potential
based on the Maxwell-Boltzmann statistics reads
\begin{align}\label{B}
\Omega_{T\neq0}& =-\frac{1}{\beta} \sum_{j,k_{z},\sigma}D_L
e^{-\beta(\epsilon^l_{jk_{z}} +\epsilon_{\sigma}^{ze} - \mu )}.
\end{align}
Then equations of the dimensionless chemical potential
$\mu^{\prime}$ and magnetization density $\overline{M}^{B}$ are
derived respectively,
\begin{align}\label{nB}
1=\overline{\omega}t^{1/2}\sum_{\sigma}\frac{e^{-\frac{1}{t}
(\frac{\overline{\omega}}{2}-g\sigma\overline{\omega}-\mu^{\prime})}}
{1-e^{-\frac{\overline{\omega}}{t}}}
\end{align}
and
\begin{align}\label{MB} \overline{M}^{B}_{T\neq 0}
=&\;t^{3/2}\sum_{\sigma}\biggl\{\frac{e^{-\frac{1}{t}(\frac{\overline{\omega}}{2}-g\sigma\overline{\omega}-\mu^{\prime})}}{1-e^{-\frac{\overline{\omega}}{t}}}\nonumber\\
&\times[1+\frac{\overline{\omega}}{t}(g\sigma-\frac{1}{2}-\frac{e^{-\frac{\overline{\omega}}{t}}}{1-e^{-\frac{\overline{\omega}}{t}}})]\biggr\}.
\end{align}
Substituting Eq.~(\ref{nB}) into Eq.~(\ref{MB}), yields
\begin{align}\label{MB1}
\overline{M}^{B}_{T\neq 0} =
\frac{1}{x^{\prime}}-\frac{1}{2}-\frac{1}{e^{x^{\prime}}-1}+\frac{g(e^{2gx^{\prime}}-1)}{e^{2gx^{\prime}}+e^{gx^{\prime}}+1}.
\end{align}
An analytical formula for $g_{c}$ can be obtained,
\begin{align}\label{MB0}
\frac{1}{2}=
\frac{1}{x^{\prime}}-\frac{1}{e^{x^{\prime}}-1}+\frac{g_c(e^{2g_cx^{\prime}}-1)}{e^{2g_cx^{\prime}}+e^{g_cx^{\prime}}+1}.
\end{align}
The value of $g_c$ can be derived from Eq.~(\ref{MB0}) exactly in
two limit cases: $g_{c}|_{t\to\infty}= 1/\sqrt{8}\approx 0.35355$
and $g_{c}|_{t\to 0}= 1/2$. The value of $g_{c}$ for a Boltzmann gas
is reasonably equal to that of a Bose gas in the high temperature
limit. Figure \ref{Gc-Bz} plots the numerical solutions of
Eq.~(\ref{MB0}) and compares with the Bose gas. An interesting point
is that the low-temperature-limit value of $g_{c}$ is also
consistent with that of a Bose gas at low temperature but in high
magnetic field, although the Maxwell-Boltzmann statistics is just
valid in the high temperature region. The stronger the field, the
better the accordance.

The $g_c-1/t$ curves in Figs.~{\ref{Gc}} and \ref{Gc-Bz} mark the
boundary between the diamagnetic and paramagnetic regions. The gas
exhibits diamagnetism at all temperatures and in all magnetic field
when $g<g_{c}|_{t\to\infty}\approx 0.35355$, while always
paramagnetism when $g\ge 0.5$. Whereas, the magnetic properties seem
more complicated in the intermediate region, $g_{c}|_{t\to\infty} <
g <0.5$.  Figures \ref{M}(a-c) illustrate the dimensionless
magnetization density $\overline{M}$ for the three different cases,
respectively.

Figure \ref{M}(a) denotes the case of $g=0.35<g_{c}|_{t\to\infty}$.
The temperature-dependence of $\overline{M}$ is very similar to that
of charged scalar Bose gases,\cite{Toms} as if the paramagnetic
effect associated with the spin degree of freedom is thoroughly
hidden. The diamagnetism is even stronger at lower temperatures. As
the external field tends to be weak, a sharp bend appears gradually
on the curve, which located at the point corresponding to the BEC
temperature in zero field.\cite{Toms} In our model, the BEC
temperature for a spin-1 gas is $(1/t)_c\approx 3.945$.
Figure~\ref{M}(c) shows $\overline{M}$ for a paramagnetic case to
the contrary when $g=0.5$. $\overline{M}$ is always positive in the
field at all temperatures. An interesting phenomenon is that the
$\overline{M}-1/t$ curve shows up a peak in this case. The decline
in $\overline{M}$ at low temperatures is attributed to the
diamagnetic effect. When weakening the magnetic field, the peak is
lowered and moves to low temperatures. Fig.~\ref{M}(b) depicts the
case with $g$ in the intermediate region, which looks quite similar
to Fig.~\ref{M}(c). The key difference is that $\overline{M}$ can
change its sign from positive to negative as the temperature
decreases, indicating that the system undergoes a shift from
paramagnetism to diamagnetism.

Figures \ref{M} demonstrate the total magnetic performance of the
charged spin-1 Bose gas. We now turn to examine the underlying
paramagnetic and diamagnetic effects for each corresponding case. As
shown in Figs.~\ref{Mt}, both the paramagnetization density and the
diamagnetization density become more stronger at lower temperatures.
As the external field is reduced, the strengthening of
$\overline{M}_p$ and $\overline{M}_d$ becomes fast near the
temperature close to the BEC point in zero field. As $g$ grows,
$\overline{M}_d$ is only slightly strengthened but $\overline{M}_p$
increases significantly and thus it can go beyond $\overline{M}_d$.
$\overline{M}_p$ thoroughly exceeds $\overline{M}_d$ when $g$ rises
to $0.5$. If $g$ increases further, the paramagnetic effect can be
so strong as to cover up the diamagnetic effect completely. Then the
total magnetization density $\overline{M}$ becomes monotonously
increasing with lowering the temperature and finally reaches a
plateau, instead of a peak, at low temperatures.

\section{Summary}

This paper has studied the interplay between paramagnetism and
diamagnetism of the ideal charged spin-1 Bose gas. The Lande-factor
$g$ is introduced to describe the strength of paramagnetic effect
caused by the spin degree of freedom. The gas exhibits a shift from
diamagnetism to paramagnetism as $g$ increases. The critical value
of $g$, $g_c$, is determined by evaluating the dimensionless
magnetization density $\overline{M}$. Our results show that $g_c$
increases monotonically as $t$ decreases. In the high temperature
limit, $g_c$ goes to a universal value in all different magnetic
fields, $g_{c}|_{t\to\infty}=1/\sqrt{8}$. At low temperatures, our
results indicate that $g_{c}$ ranges from $0.475$ to $0.50$ as the
magnetic field varies. Therefore, a gas with $g<{1/\sqrt{8}}$
exhibits diamagnetism at all temperatures, but a gas with $g>{1/2}$
always exhibits paramagnetism. In cases where ${1/\sqrt{8}}<g<1/2$,
the Bose gas undergoes a shift from paramagnetism to diamagnetism as
the temperature decreases.

In order to depict some details of the competition between
paramagnetism and diamagnetism, the paramagnetic and diamagnetic
contributions to the total magnetization density are also
calculated. No doubt that the paramagnetism is enhanced with
increasing $g$. Surprisingly, the diamagnetism is not suppressed,
but slightly strengthen. This implies that the competition between
para- and dia-magnetism is nontrivial. When $g$ is fixed, both the
paramagnetism and diamagnetism become stronger as $t$ decreases. As
in the scalar case, there is no Bose-Einstein condensation when the
magnetic field is present, no matter how small it is. However,
evidence of the condensation can be seen in the magnetization
density as the magnetic field is reduced.

At last, we briefly discuss experimental aspects possibly relevant
to the present work. Although the charged spinor Bose gas has not
been realized so far, the up-to-date achievement in experiments
makes it attainable perhaps in the near future.\cite{Kabanov2005}
For example, it is already possible to create ultracold plasmas by
photoionization of laser-cooled neutron atoms.\cite{Killian1999} The
ions can be regarded as charged bosons if their spin is an integer.
The Lande-factor for different magnetic ions could be different. As
reported by Killian {\it et al.}\cite{Killian1999}, the temperatures
of electrons and ions are as low as 100 mK and 10 $\mu$K,
respectively. One can expect that the temperature could be lowered
near to the BEC temperature with new advancement in experimental
techniques. The diamagnetism-paramagnetism effects manifest
themselves near the BEC point. On the other hand, several
ferromagnetic superconductors have been discovered since
2000.\cite{Saxena2000} Cooper pairs in such materials are likely in
the spin-triplet state, thus behave somewhat like charged spin-1
bosons. Note that the charged spin-1 boson model can not be applied
directly to describe triplet superconductors, but it could offer
some help in understanding magnetic properties of such materials.

This work was supported by the Fok Ying Tung Education Foundation of
China (No.~101008), the Key Project of the Chinese Ministry of
Education (No.~109011), and the Fundamental Research Funds for the
Central Universities of China.

\end{document}